\documentclass[12pt,manuscript]{aastex63}

\usepackage{graphicx}
\usepackage{longtable}
\usepackage{hyperref}
\def \fv {}
\def \fvv{}   
\def \cc {}
\def\agile{{\fv{\it AGILE}} }
\def\agilep{{\fv{\it AGILE}}}
\def \chf{CHIME/FRB }
\def \chfp{CHIME/FRB}
\def \sgr{SGR 1806-20 }
\def \sgrp{SGR 1806-20}
\def \frbp{FRB 180916}
\def \frb{FRB 180916 }

\def \ffrb{FRB 180916.J0158+65 }
\def\swift{{\it Swift} }
\def\swiftp{{\it Swift}}

\def\mt{}
\def\mtt{}

\def\be{\begin{equation}}
\def\en{\end{equation}}

\begin{document}
\title{\large{Gamma-Ray and X-Ray Observations of the Periodic-Repeater FRB~180916 During Active Phases}}
\author{M. Tavani}
\email{marco.tavani@inaf.it}
\affiliation{\scriptsize INAF/IAPS, via del Fosso del Cavaliere 100, I-00133 Roma (RM), Italy}
\affiliation{\scriptsize Universit\`a degli Studi di Roma Tor Vergata, via della Ricerca Scientifica 1, I-00133 Roma (RM), Italy}

\author{F. Verrecchia}
\email{francesco.verrecchia@inaf.it}
\affiliation{\scriptsize SSDC/ASI, via del Politecnico snc, I-00133 Roma (RM), Italy}
\affiliation{\scriptsize INAF/OAR, via Frascati 33, I-00078 Monte Porzio Catone (RM), Italy}

\author{C. Casentini}
\email{claudio.casentini@inaf.it}
\affiliation{\scriptsize INAF/IAPS, via del Fosso del Cavaliere 100, I-00133 Roma (RM), Italy}
\affiliation{\scriptsize INFN Sezione di Roma 2, via della Ricerca Scientifica 1, I-00133 Roma (RM), Italy}

\author{M. Perri}
\affiliation{\scriptsize SSDC/ASI, via del Politecnico snc, I-00133 Roma (RM), Italy}
\affiliation{\scriptsize INAF/OAR, via Frascati 33, I-00078 Monte Porzio Catone (RM), Italy}

\author{A. Ursi}
\affiliation{\scriptsize INAF/IAPS, via del Fosso del Cavaliere 100, I-00133 Roma (RM), Italy}

\author{L. Pacciani}
\affiliation{\scriptsize INAF/IAPS, via del Fosso del Cavaliere 100, I-00133 Roma (RM), Italy}

\author{C. Pittori}
\affiliation{\scriptsize SSDC/ASI, via del Politecnico snc, I-00133 Roma (RM), Italy}
\affiliation{\scriptsize INAF/OAR, via Frascati 33, I-00078 Monte Porzio Catone (RM), Italy}

\author{A. Bulgarelli}
\affiliation{\scriptsize INAF/OAS, via Gobetti 101, I-40129 Bologna (BO), Italy}

\author{G. Piano}
\affiliation{\scriptsize INAF/IAPS, via del Fosso del Cavaliere 100, I-00133 Roma (RM), Italy}

\author{M. Pilia}
\affiliation{\scriptsize {\fv {INAF/OACa, via della Scienza 5, I-09047 Selargius (CA), Italy}}}

\author{G. Bernardi}
\affiliation{\scriptsize INAF/IRA, via Piero Gobetti 101, 40129 Bologna (BO), Italy}
\affiliation{\scriptsize Department of Physics and Electronics, Rhodes University, PO Box 94, Grahamstown, 6140, South Africa}

\author{A.~Addis}
\affiliation{\scriptsize INAF/OAS, via Gobetti 101, I-40129 Bologna (BO), Italy}

\author{L.A. Antonelli}
\affiliation{\scriptsize INAF/OAR, via Frascati 33, I-00078 Monte Porzio Catone (RM), Italy}
															
\author{A. Argan}
\affiliation{\scriptsize INAF/IAPS, via del Fosso del Cavaliere 100, I-00133 Roma (RM), Italy}
\author{L.~Baroncelli}
\affiliation{\scriptsize INAF/OAS, via Gobetti 101, I-40129 Bologna (BO), Italy}
\affiliation{\scriptsize Dip. di Fisica e Astronomia, Universit\`a di Bologna, Viale Berti Pichat 6/2, 40127, Bologna, Italy}

\author{P. Caraveo}
\affiliation{\scriptsize INAF/IASF, via E. Bassini 15, I-20133 Milano (MI), Italy}
\affiliation{\scriptsize INFN Sezione di Pavia, via U. Bassi 6, I-27100 Pavia (PV), Italy}

\author{P.W. Cattaneo}
\affiliation{\scriptsize INFN Sezione di Pavia, via U. Bassi 6, I-27100 Pavia (PV), Italy}
\author{A. Chen}
\affiliation{\scriptsize School of Physics, Wits University, Johannesburg, South Africa}

\author{E. Costa}
\affiliation{\scriptsize INAF/IAPS, via del Fosso del Cavaliere 100, I-00133 Roma (RM), Italy}

\author{G. Di Persio}
\affiliation{\scriptsize INAF/IAPS, via del Fosso del Cavaliere 100, I-00133 Roma (RM), Italy}

\author{I. Donnarumma}
\affiliation{\scriptsize ASI, via del Politecnico snc, I-00133 Roma (RM), Italy}

\author{Y. Evangelista}
\affiliation{\scriptsize INAF/IAPS, via del Fosso del Cavaliere 100, I-00133 Roma (RM), Italy}

\author{M. Feroci}
\affiliation{\scriptsize INAF/IAPS, via del Fosso del Cavaliere 100, I-00133 Roma (RM), Italy}

\author{A. Ferrari}
\affiliation{\scriptsize CIFS, c/o Physics Department, University of Turin, via P. Giuria 1, I-10125,  Torino, Italy}

\author{V.~Fioretti}
\affiliation{\scriptsize INAF/OAS, via Gobetti 101, I-40129 Bologna (BO), Italy}

\author{F. Lazzarotto}
\affiliation{\scriptsize INAF/OAPd, vicolo Osservatorio 5, I-35122 Padova (PD), Italy}

\author{F. Longo}
\affiliation{\scriptsize Dipartimento di Fisica, Universit\`a di Trieste and INFN, via Valerio 2, I-34127 Trieste (TR), Italy}

\author{A. Morselli}
\affiliation{\scriptsize INFN Sezione di Roma 2, via della Ricerca Scientifica 1, I-00133 Roma (RM), Italy}
\author{F. Paoletti}
\affiliation{\scriptsize East Windsor RSD, 25A Leshin Lane, Hightstown, NJ-08520, USA}
\affiliation{\scriptsize INAF/IAPS, via del Fosso del Cavaliere 100, I-00133 Roma (RM), Italy}
\author{N.~Parmiggiani}
\affiliation{\scriptsize INAF/OAS, via Gobetti 101, I-40129 Bologna (BO), Italy}
\affiliation{\scriptsize Universit\`a degli Studi di Modena e Reggio Emilia, DIEF - Via Pietro Vivarelli 10, 41125 Modena, Italy}
\author{A. Trois}
\affiliation{\scriptsize {\fv {INAF/OACa, via della Scienza 5, I-09047 Selargius (CA), Italy}}}
\author{S.~Vercellone}
\affiliation{\scriptsize INAF/OAB, via Emilio Bianchi 46, I-23807 Merate (LC), Italy}

\author{G. Naldi}
\affiliation{\scriptsize INAF/IRA, via Piero Gobetti 101, 40129 Bologna (BO), Italy}
\author{G. Pupillo}
\affiliation{\scriptsize INAF/IRA, via Piero Gobetti 101, 40129 Bologna (BO), Italy}
\author{G. Bianchi}
\affiliation{\scriptsize INAF/IRA, via Piero Gobetti 101, 40129 Bologna (BO), Italy}

\author{S. Puccetti}
\affiliation{\scriptsize ASI, via del Politecnico snc, I-00133 Roma (RM), Italy}


\submitjournal{the {\it Astrophysical Journal Letters}: March 11, 2020; accepted: April 5, 2020.}

\shorttitle{\small $\gamma$-Ray and X-Ray Observations of the Repeater FRB 180916 During Active Phases}

\shortauthors{\small Tavani et al.}

\begin{abstract}

\frb is a most intriguing source capable of producing repeating fast
radio bursts with a {\mt periodic} 16.3 day temporal pattern.
The source is well positioned in a star forming region in the
outskirts of a nearby galaxy at 150 Mpc distance. In this paper
we report on the X-ray and $\gamma$-ray observations of \frb
obtained by \agile and \swiftp. We focused especially on {\mt
the recurrent 5-day time intervals of enhanced radio bursting}.
{\mt In particular, we report on}
the results
obtained in the time intervals {\mtt {Feb.\,3\,-\,8; Feb. 25; Mar.\,5\,-\,10; Mar.\,22\,-\, 28, 2020}}
during a multiwavelength campaign {\mt involving high-energy
and radio observations of \frbp}. We also searched for temporal
coincidences at millisecond timescales between the 32 known radio
bursts of \frb and X-ray and MeV events detectable by \agilep.
We do not detect any simultaneous event or any extended X-ray
and $\gamma$-ray emission on timescales of hours/days/weeks.
Our {\mtt cumulative} X-ray {\mtt (0.3-10 keV)} flux upper limit
of $5 \times\,10^{-14}\, \rm \, erg \,cm^{-2} \, s^{-1} $
{\mtt (obtained during 5-day active intervals from several 1-2 ks integrations)}
translates into an isotropic luminosity upper limit
of $L_{X,UL} \sim 1.5 \times\, 10^{41} \, \rm erg \, s^{-1}$.
Deep $\gamma$-ray observations above {\fvv {100}} MeV over a many-year
timescale provide an average luminosity upper limit one order of magnitude larger.
These results provide the so-far most stringent upper limits on high-energy
emission from the \frb source. {\cc{Our results constrain the dissipation}} of magnetic energy from a magnetar-like source of {\mt radius $R_m$}, internal magnetic field $B_m$ and dissipation
timescale $\tau_d$
to satisfy the relation $R_{m,6}^3 \, B_{m,16}^2 \, \tau_{d,8}^{-1} \lesssim 1$, where {\mt $R_{m,6}$ is $R_m$ in units of $10^6$ cm}, $B_{m,16}$ is $B_m$ in units of $10^{16}$ G, and $\tau_{d,8}$ in units of $10^8$ {\cc {s}}.

\end{abstract}

\section{Introduction}

Fast Radio Bursts (FRBs) are impulsive outbursts of mysterious origin (typically lasting $\sim$ 1 msec) detected in the sub-GHz and GHz bands  \citep[e.g.,][]{2007Science.218.777,2019ARA&A119..161101}. Most of the FRBs have been detected only once, but a special category of repeating FRBs has been recently revealed, demonstrating that a sub-class of sources can produce multiple radio outbursts over timescales of years \citep[][]{2019ARA&A119..161101,2019AeARv27..4P}.

We focus here on the remarkable repeater \ffrb (hereafter \frbp) for which {\mt {\fvv {35}} radio bursts} have been detected {\mt with a periodic pattern} during the {\fv {time interval}} Sept. 2018 - {\mt Oct.} 2019
{\mt \citep[][hereafter C20]{2020ApJLCHIME}}.
This source, originally localized by CHIME/FRB with an uncertainty of
about 0.09 square degrees \citep[][hereafter C19]{2019ApJLCHIMEc} was
later detected by VLBI observations
a precise arcsec positioning
\citep[][hereafter M20]{2020Natur577...190}.
The source is positionally coincident with the outskirts of a spiral galaxy at the distance of about 150 Mpc. The 3-sigma significance of the proximity of \frb with a galaxy at a redshift $z\, =\, 0.0337\, \pm\, 0.0002$ and the a-posteriori determination of the intergalactic contribution to its observed excess dispersion measure (DM)
strongly support the association of \frb with {\cc {the galaxy SDSS J015800.28+654253.0 as reported in M20}}. This repeating FRB source is therefore located at a distance {\mt significantly} smaller than {\mt that of } the majority of FRBs showing $ DM_{excess} \,\geq\, 100 \rm \, pc\,cm^{-3}$ {\fv {where $DM_{excess}$ is the excess DM once the Galactic and halo contributions are taken into account}}.

Evidence of periodic activity of radio bursting from \frb  provides important clues on the mechanism of FRB generation from this source {\mt (C20)}. Radio bursts are not randomly distributed but appear to concentrate especially in specific 5-day intervals recurring every
16.35 days. {\mtt As noted in C20, this period can be affected by aliasing\footnote{\mtt The resulting period could be in the range of a few hours to a day as discussed in C20. However, C20 discuss several possibilities that support the hypothesis that this is not the case.} due to the
\chf regular exposure pattern of \frbp.} A periodic behavior of an otherwise unpredictable emission is
remarkable, and 
suggests {\mt either} a "triggering" of the radio bursting {\cc {from}} physical interactions in a binary system, {\mt or emission influenced by compact star precession}.

\frb is then worth a careful study, given its relevance in terms of proximity, number of detected {\mt radio bursts}, and periodic activity. Table~\ref{tab:tab1} summarizes the relevant information on the 32 radio bursts so far detected from \frbp, providing
{\mt topocentric} arrival times and radio fluences on millisecond timescales. The (isotropic-equivalent) energy involved in the radio burst emissions is {\mt in the range} ${E_{radio, iso}}\, \simeq \, (10^{36} - 10^{38} {\rm erg})$ at the source distance, {\mt that is a} value million times larger than {\mt that of} the brightest giant pulses observed from the Crab pulsar \citep[][]{2003Natur422...141H}. The object {\mt underlying \frb} is therefore exceptional in its emission properties.

We recently presented in \citet[][]{2020Casentini} preliminary results of \agile observations of \frb based on the first 10 detected radio bursts. In this paper we extend our search for high-energy emission from the source considering \swift X-ray observations as well as \agile $\gamma$-ray data both focused on specific active 5-day repetition intervals. Our observations are part of a multi-frequency campaign involving \agilep, Swift and radio observations of \frb during the period{\fv {s}} {\cc {Feb.\,3 -- Feb.\,8}} {\fv {and Feb.\,19 -- Feb.\,25}}, 2020 carried out at the Sardinia Radio Telescope \citep[][]{2017PrandoniSRT} and at the Northern Cross in Medicina \citep[][]{2020NCSubmitted}. Our improved upper limits on X-ray and $\gamma$-ray emissions from \frb provide {\cc {significant}} constraints on the physical nature of \frbp. \\

\section{\agile Observations}

The \agile mission is an ASI (Italian Space Agency) space project dedicated to X and $\gamma$-ray astrophysics  \citep[][]{2009A&A...502..995T,2019RLSFN...38}. The instrument consists of four different detectors onboard the satellite: an imaging $\gamma$--ray Silicon Tracker sensitive in the range 30 MeV - 30 GeV \citep[GRID;][]{2002NIMPA.490..146B}, a {\fv {coded mask}} X-ray imager sensitive in the 18-60 keV band \citep[Super-\agilep;][]{2007NIMA581..728}, the Mini-Calorimeter sensitive in the 0.4-100 MeV band \citep[MCAL;][]{2009NIMPA.598..470L} and the anti-coincidence \citep[][]{2006NIMPA.556..228P} system.
Currently, \agile operates in spinning mode \citep[][]{PittoriSSDC}, with the instrument axis rotating every $\sim 7$ minutes around the {\mt satellite-sun direction}; for each satellite revolution, an exposure of about 80\% of the entire sky is obtained. For a summary of the \agile mission features, see \citet[][]{2019RLSFN...38}.

\agile is capable of quasi-continuously monitoring a source positioned in the accessible portion of the sky. Source visibility depends on solar panel constraints (a seasonal effect that makes about 20\% of the sky near the Sun or in the opposite direction not available for observations at any given time), Earth occultations (affecting source visibility with $\sim 10^3$ {\cc {s}} timescales), and South Atlantic Anomaly (affecting about 10\% of the {\cc {$\sim$\,95-min}} satellite orbit). Despite these limitations, \agile can be very effective in detecting transient and steady X-ray and $\gamma$-ray emissions from cosmic sources.

In this paper we focus on \agile and \swift observations of the \frb location (Galactic coordinates $l = {\cc {129.7^\circ}}$, $b = {\cc {3.7^\circ}}$) that turns out to be somewhat influenced by diffuse Galactic emission. {\fv {The results of the analysis of radio observations \citep[][]{2020ATel13492} will be reported in a separated paper \citep[][]{2020piliaToBeSub}}}. In the following we report the results of searches for high energy emission in coincidence with the arrival times (T$_{0}$) of Table \ref{tab:tab1} within time windows of 200 seconds centered at T$_{0}$.
{\mtt Due to dispersive delay, radio waves reach detectors on Earth or in orbit several seconds after X-rays or $\gamma$-rays. In our searches we take as $T_0$'s the de-dispersed topocentric arrival times\footnote{The difference between the actual AGILE arrival time and the radiotelescopes' de-dispersed topocentric times is of order of 20-40 msec, and it is negligible for our adopted window search of 200 sec.} at infinite frequencies for the different radio telescopes of Table \ref{tab:tab1}. }

\begin{table*}[]
\scriptsize
    \caption{FRB180916.J0158+65 bursts and \agile observations} \vskip1cm
  \begin{center}
\vskip-1.0cm
    \begin{tabular}{|c@{\hspace{5mm}}c@{\hspace{6mm}}c@{\hspace{8mm}}c@{\hspace{4mm}}c@{\hspace{4mm}}c@{\hspace{4mm}}|c@{\hspace{4mm}}c@{\hspace{4mm}}c@{\hspace{4mm}}|} 
     \hline
     \multicolumn{6} {|c@{\hspace{5mm}}} {FRB parameters} & \multicolumn{3} {c|} {Event in the \agile FoV}\\
     \hline
    Burst no. & Day & Arrival times{\tablenotemark{\scriptsize{*}}} & Instr.{\tablenotemark{{\fvv{\scriptsize{+}}}}} & bandwidth & Fluence & MCAL & GRID & Super-A\\
    & (yymmdd) & (UTC) & & (MHz) & (Jy ms) & & & \\
     \hline
       1 & $180916$ & 10:15:10.746748 & C & 400-800  & $>$2.3 & NO & NO & NO\\
       2 & $181019$ & 08:13:13.695545 & C & 400-800  & $>$3.5 & - & - & -\\
       3 & $181019$ & 08:13:13.759080 & C & 400-800  & $>$2.0 & - & - & -\\
       4 & $181104$ & 06:57:09.522136 & C & 400-800  & $>$2.8 & YES & NO & NO\\
       5 & $181104$ & 07:06:52.655909 & C & 400-800  & 6.8 & YES & NO & NO\\
       6 & $181120$ & 05:55:57.158955 & C & 400-800  & 8.0 & YES & NO & NO\\
       7 & $181222$ & 03:59:14.238510 & C & 400-800  & 9.6/15/16.5/7.2 & YES & YES & NO\\
       8 & $181223$ & 03:51:19.892113 & C & 400-800  & 10.4/3.6 & YES & YES & NO\\
       9 & $181225$ & 03:52:54.884107 & C & 400-800  & 3.1 & YES & NO & NO\\
     10  & $181226$ & 03:43:21.063337 & C & 400-800  & 2.9/1.6 & YES & NO & NO\\
     11  & $190126$ & 01:32:41.297440 & C & 400-800  & 6.4 & - & - & -  \\
     12  & $190518$ & 18:13:33.601287 & C & 400-800  & 1.0 & - & - & - \\  
     13  & $190518$ & 18:20:56.937173 & C & 400-800  & 7.7 & NO & NO & NO  \\
     14  & $190519$ & 17:50:16.601726 & C & 400-800  & $>$1.3/$>$2.2 & YES & YES & NO  \\
     15  & $190519$ & 18:08:52.454119 & C & 400-800  & 3.1/5.3 & - & - & -  \\
     16  & $190519$ & 18:10:41.139009 & C & 400-800  & 4.4 & NO & NO & NO  \\
     17  & $190603$ & 17:09:14.367599 & C & 400-800  & 2.1 & YES & NO & NO  \\
     18  & $190604$ & 17:11:30.757683 & C & 400-800  & 37 & - & - & - \\ 
     19  & $190605$ & 16:55:55.866635 & C & 400-800  & $>$7.0 & NO & NO & NO  \\
     20  & $190605$ & 17:02:21.845335 & C & 400-800  & 11.5/5.4 & NO & NO & NO  \\
     21  & $190619$ &  02:24:19.980000 & E  & 1636-1764 & 0.72 & YES & NO & NO  \\ 
     22  & $190619$ &  02:46:06.367000 & E  & 1636-1764 & 0.20 & YES & YES & NO \\ 
     23  & $190619$ &  03:36:59.608000 & E  & 1636-1764 & 0.62 & - & - & -  \\ 
     24  & $190619$ &  06:46:56.344000 & E  & 1636-1764 & 2.53 & YES & NO & NO  \\ 
     25  & $190809$ & 12:50:40.150674 & C & 400-800 & 7.3 & YES & NO & NO  \\
     26  & $190810$ & 12:49:34.408294 & C & 400-800 & $>$1.7 & - & - & - \\  
     27  & $190825$ & 11:48:18.617542 & C & 400-800 & 24 & NO & NO & NO  \\
     28  & $190825$ & 11:51:53.904598 & C & 400-800 & 2.8 & NO & NO & NO  \\
     29  & $190825$ & 11:51:53.967298 & C & 400-800 & 1.4 & NO & NO & NO  \\
     30  & $190825$ & 11:53:35.794063 & C & 400-800 & $>$2.6 & NO & NO & NO  \\
     31  & $191030$ & 07:33:47.920217 & C & 400-800 & 2.3 & - & - & - \\  
     32  & $191030$ & 07:41:43.668377 & C & 400-800 & $>$2.3 & YES & YES & YES  \\

     33  & $200220$ & 13:27:35.57 & S & 296-360 & 37 & YES & YES & -  \\
     34  & $200220$ & 13:36:49.57 & S & 296-360 & 13 & - & - & -  \\
     35  & $200220$ & 13:36:49.57 & S & 296-360 & 19 & - & - & -  \\

   \hline
    \end{tabular}
    \label{tab:tab1}
  \end{center}
\vskip-.1cm
Notes:\\
{*}: De-dispersed topocentric arrival times at the CHIME{\fvv{ , Effelsberg or SRT}} locations.\\
{\fvv {+: Radio instruments which detected each specific burst, where C is for \chfp, E for EVN and S for SRT.\\
Unmarked detector FoV labels are
for radio events occurring when the AGILE instrument was in idle mode with no data acquisition mostly due to passages over the SAA.}}
\end{table*}

\subsection{Super-A flux upper limits}

Super-AGILE detects steady and transient emission from sources in the 18-60 keV energy
band within a field of view of $107{\cc{^\circ}}\,\times\, 68{\cc{^\circ}}$.
Due to the spinning of the AGILE satellite, we 
{\mt analyzed} time periods with the source
within 10$^{\circ}$ from the {\mt instrument} on-axis direction.
The 
{\mt resulting} time windows have a duration of about 10-20 s.
We excluded events taken during the passage above the South Atlantic Anomaly, and events collected when the source was occulted by the Earth.
We {\mt searched for} hard X-ray emission during the 5-day active periods of \frb
during the intervals {\fvv { Feb.}} 3\,$-$\,8, 2020, and {\fvv { Feb.}} 19 $-$ 25, 2020.
The corresponding effective exposures of 5.6 and 2.4 ks 
{\mt lead to} 5$\sigma$ ULs of
2.4 and 5.0 $\times$10$^{-9}$ erg cm$^{-2}$ s$^{-1}$ for the two {\mt
observing intervals}.
For 
 {\mt the sum of observations during the interval} Feb. 3\,$-$\, 25, 2020, we obtain an upper limit of 1.8$\times$10$^{-9}$ erg\, cm$^{-2}$\, s$^{-1}$ for a total exposure of 13.5 ks.

\subsection{MCAL fluence upper limits}

The \agilep/{\fv {MCAL}} \citep{2009NIMPA.598..470L,2008A&A...490.1151M} is an all-sky monitor with a $4\pi$ {\fv {FoV}}, sensitive in the 400~keV~--~100~MeV energy range. It is a segmented detector, composed of 30 CsI(Tl) scintillation bars, with a total geometrical area of $\sim1400$~cm$^{2}$ (on-axis). The \agile MCAL is a triggered detector: its onboard trigger logic acts on different energy ranges and timescales (ranging from $\sim300\, \mu$s to $\sim8$~s). The sub-millisecond logic timescale
 {\mt allows MCAL to trigger on} very short-duration impulsive events. {\mt MCAL is
  then capable of efficiently detecting long and short transients such as} GRBs \citep{Galli2013} as well as Terrestrial Gamma-ray Flashes (TGFs) {\mt on millisecond timescales} \citep{Marisaldi2015}. A detailed discussion about MCAL triggering and upper limit capabilities {\mt is reported} in \citep[][]{2020Casentini,2019ApJ...871...27U}.

We checked the possible detections by MCAL of transient emission at or near the times of arrival of all accessible radio burst events marked with a "YES" label in the MCAL column of Table \ref{tab:tab3}. Our search was within $\pm 100$ seconds from the topocentric arrival times of Table \ref{tab:tab1}.

No significant event was detected for a typical fluence UL obtained for millisecond timescales  of $10^{-8} \rm \, erg \, cm^{-2}$. Table \ref{tab:tab3} provides the typical MCAL fluence ULs obtained at different {\mt trigger} timescales. When compared with the radio fluences of \frb bursts, our ULs in the MeV range set the constraint that the energy detectable in the MeV range and emitted on millisecond timescales have to be less than $\sim 10^8$ times the energy emitted in the GHz band ($E_r \sim 3 \,\times\, 10^{37} \rm \, erg$).

\begin{table*}[ht!]
  \begin{center}
    \caption{Average \agilep/{\fv {MCAL}} fluence ULs in {\cc {$\rm erg \, cm^{-2}$}}}
    \begin{tabular}{|c|c|c|c|c|c|c|}
     \hline
    sub-ms & 1 ms & 16 ms & 64 ms & 256 ms & 1024 ms & 8192 ms\\
     \hline

     $1.13 \times 10^{-8}$ & $1.29 \times 10^{-8}$ & $3.72 \times 10^{-8}$ & $4.97 \times 10^{-8}$ & $7.95 \times 10^{-8}$ & $1.59 \times 10^{-7}$ & $4.49 \times 10^{-7}$\\

     \hline
    \end{tabular}
    \label{tab:tab3}
 \end{center}
\end{table*}

\subsection{GRID flux upper limits}

Gamma-ray observations of \frb focused on day-long exposures of the source during the
 {\mtt reportedly} active time intervals following the 16.3 periodicity of C20. Table \ref{tab:tab4} summarizes the \agilep/GRID ULs obtained above 100 MeV for 5-day integrations corresponding to 10 contiguous active intervals of \frb covering the period 25 Aug. 2019\, --\, 28 Mar. 2020. Typical 5-day GRID flux ULs are $10^{-10} \rm \, erg \, cm^{-2} \, s^{-1}$ {\fvv (obtained by the AGILE multi-source maximum likelihood analysis for a power-law spectral model of spectral photon index of 2)} corresponding to isotropic-equivalent $\gamma$-ray luminosity ULs of
$3 \times\, 10^{44} \rm \, erg \, s^{-1}$.

In addition to the latter integrations, we also performed a deep exposure of the \frb region including both pointing (2007\,--\,2009) and spinning (2010\,--\,today) mode data.
We find an integrated flux UL $ \sim 8.2 \times\, 10^{-13} \, \rm erg \, cm^{-2} \, s^{-1}$, corresponding to an isotropic long-term averaged $\gamma$-ray luminosity $L_{\gamma,UL,ave} \sim 2 \times\, 10^{42} \, \rm erg \, s^{-1}$.

\begin{table*}[ht!]
    \caption{Average \agilep/{\fv {GRID}} flux ULs in {\cc {$\rm erg \, cm^{-2} \, s^{-1}$}}.}
    \begin{tabular}{|c|c|c|c|c|c|}
     \hline
     $\Delta T=10^1$ s& $\Delta T=10^2$ s & $\Delta T=10^3$ s & $\Delta T=10^0$ d & $\Delta T=10^1$ d& $\Delta T=10^2$ d\\
     \hline
     4.0$\times 10^{-7}$ & 3.5$\times 10^{-8}$ & 1.5$\times 10^{-8}$ & 4.0$\times 10^{-10}$ & 1.0$\times 10^{-10}$ & {\fvv {1.7\,}}$\times 10^{-11}$\\
     \hline
    \end{tabular}
    \label{tab:tab4}
\vskip-.1cm
Notes:\\
\vskip-.4cm
{2\,$\sigma$ flux ULs ($\rm erg \, cm^{-2} \, s^{-1}$) obtained for emission in the range 50\,MeV\,--\,10 GeV for the short integration timescales and 100\,MeV\,--\,10 GeV for the long ones,
{\mtt at the \frb position.}}\\
\end{table*}

\begin{table*}[ht!]
  \begin{center}
    \caption{\agilep/{\fv {GRID}} flux upper limits on 5-day integrations centered on single activity cycles. Flux integrations are centered on radio burst activity cycles derived from C20.}
    \begin{tabular}{|c|c|c|c|}
     \hline
     Cycle no. & Date{\tablenotemark{\scriptsize{*}}}  &  UL value\\
     & (UTC) & ($\times 10^{-10}\,\rm erg \, cm^{-2} \, s^{-1}$)\\
     \hline

      3  & 25/03/2020 09:00  & 2.06 \\
      2  & 09/03/2020 00:36  & 1.42 \\
      1  & 21/02/2020 16:12  & 1.36 \\
      0  & 05/02/2020 07:48  & 0.90 \\
     -1  & 19/01/2020 23:24  & 2.30 \\
     -2  & 03/01/2020 15:00  & 0.74 \\
     -3  & 18/12/2019 06:36  & 0.89\\
     -4  & 01/12/2019 22:12  & 1.00\\
     -5  & 15/11/2019 13:48  & 2.10\\
     -6  & 30/10/2019 05:24  & 1.20\\
     -7  & 13/10/2019 21:00  & 1.00\\
     -8  & 27/09/2019 12:36  & 0.76\\
     -9  & 11/09/2019 04:12  & 1.50\\
     -10 & 25/08/2019 19:48  & 1.10\\

     \hline
    \end{tabular}
    \label{tab:tab5}
 \end{center}
\vskip-.1cm
Notes:\\
\vskip-.4cm
*: Dates refer to the centers of the 5-day time intervals.\\
\end{table*}
\begin{figure}[tbp]
\begin{center}
  \centerline{\includegraphics[width=1.0\textwidth]{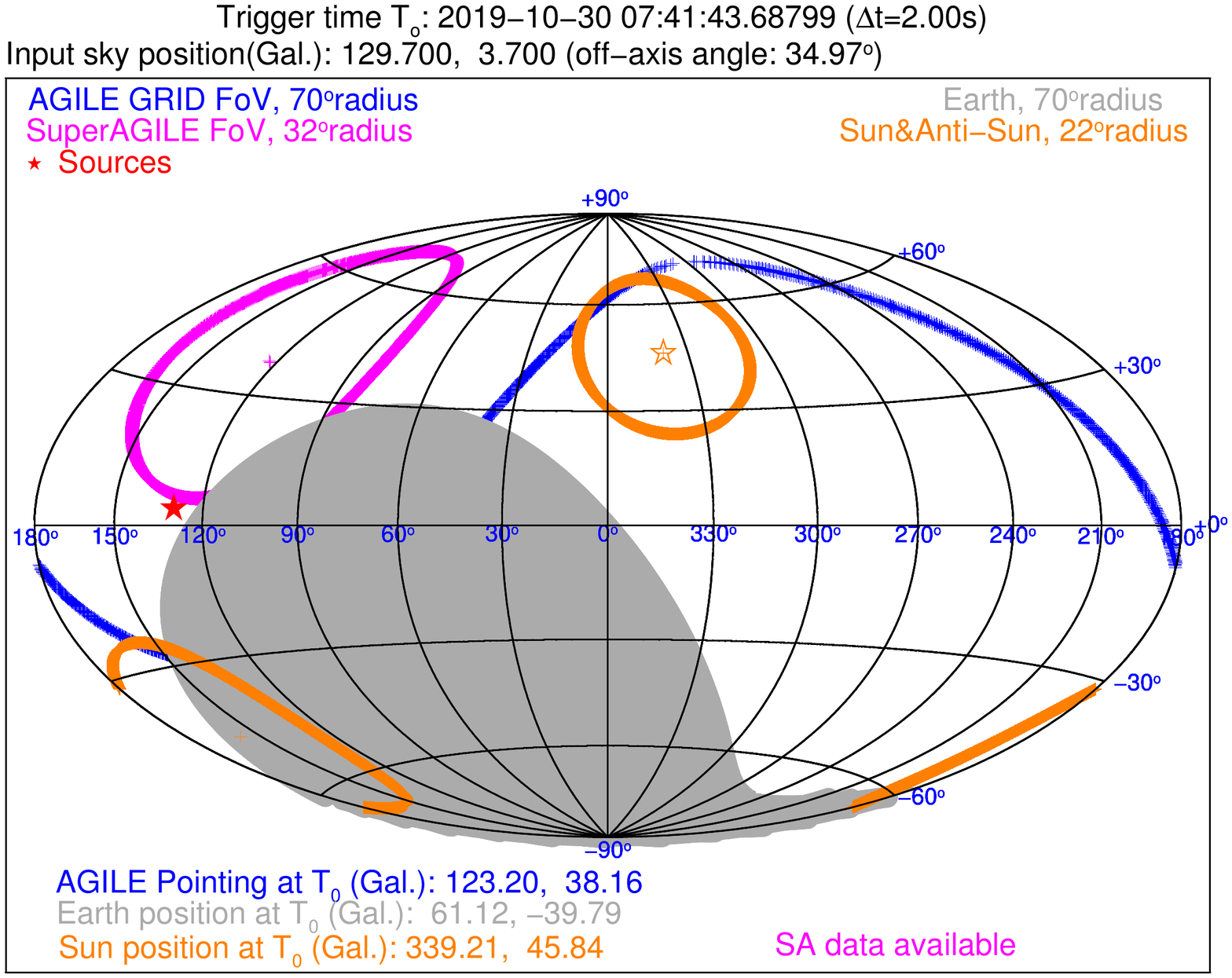}}
  \caption{{\mt Galactic coordinate map showing the {\agilep/GRID} FoV (in blue color) and the Super-A FoV (in magenta color) at the time of arrival $T_0$ of the \frb radio burst {\cc {no.\,32}} of Table \ref{tab:tab1} on {\cc {Oct.\,30}} 2019 07:41:43.668 UTC. The position of \frb is marked with a red star. The shaded gray area shows the portion of the sky occulted by the Earth at $T_0$.}   }
  \label{fig1}
\end{center}
\end{figure}

\section{\swift Observations}

The Neil Gehrels Swift Gamma-ray Burst Observatory has three instruments
\citep[][]{2004ApJ...611.1005G}; namely, the Burst Alert Telescope
\citep[BAT,][]{2005MakwardtBAT}, the X-ray Telescope \citep[XRT,][]{2005ApJL..890..L32B},
and the Ultraviolet/Optical Telescope (UVOT, Roming et al. 2005). A request for \swift
ToO observations of \frb was granted, and a {\mtt first} 6-day period of pointings started on {\cc {Feb.\,3}}, 2020. {\mt An additional observation on }{\cc {Feb.\,25, 2020}} {\mt was also carried out} {\mtt (solar panel constraints did not allow \swift to monitor the corresponding active time interval). Additional \swift data were collected during the active time intervals  Mar.\,5\, --\, 10 and Mar.\, 22\, --\,28, 2020}.

\newpage
\subsection {\swiftp/XRT flux upper limits}

The XRT data were taken within the framework of the planned coordinate multi-instrument campaign.
The \frb position was monitored in the X-rays {\mtt (0.3--10 keV) during a first interval }
with daily observations between MJD 58882 and MJD 58887 \citep[][]{2020ATel13446,2020ATel13462}. {\mtt  A single observation was carried out}
on MJD 58904. {\mtt Additional pointings were obtained during the time windows  MJD 58913\, --\, 58918, and MJD 58921\, --\, 58926}.
All observations were carried out in the Photon Counting (PC) readout mode, with a
{\mtt typical}
exposure of $\sim\, 1.5\, \rm ks$ per pointing. The data were processed using
the XRTDAS software package (v.3.5.0), developed
by the ASI Space Science Data Center (SSDC) and released by HEASARC in
the HEASoft package (v.6.26.1). The data were calibrated and
cleaned with standard filtering criteria using the xrtpipeline task
and the calibration files available from the Swift/XRT CALDB (version 20190910).
The imaging analysis was executed selecting events in the energy
channels between 0.3 keV and 10 keV and within a 10-pixel ($\sim\,$46 arcsecond) radius, which contains 80\% of the
point spread function (PSF). The background was estimated
from an annular region with radii of 25 and 60 pixels.
{\fv {No X-ray source was detected.}}
We extracted the 3-$\sigma$ countrates UL using the XIMAGE package (${\it sosta}$ {\fvv {command}})
and converted to fluxes using a
{\mtt standard single} power-law spectral model with photon index $\alpha = 2.0$ {\mtt (with the photon number flux defined as $dN/d\varepsilon \propto \varepsilon^{- \alpha}$, with $\varepsilon$ the photon energy)}. {\mtt We} corrected {\mtt the flux}
for absorption with a neutral-hydrogen {\fvv {(N$_H$)}} column density fixed to
the Galactic 21 cm value in the direction of the source, 7.1\,$\times\,10^{21}\,\rm cm^{-2}$ \citep[][]{2016HI4PI},
and redshifted {\fvv {the single spectral component}} for the known value
$z\,=\,0.0337$ (M20)\footnote{\fvv We verified that variations in
photon index of $\pm\,0.5$ leads to flux values within -20\%/+30\% with respect
to the one reported above; variations of  N$_H$
from $10^{21}$ to $10^{22}\,\rm cm^{-2}$ correspond
to variations in flux values within $\lesssim\,$30\%.}.
Combining all the {\fvv {\swiftp/}}XRT observations we obtain an overall 3-sigma {\fv {flux}} upper limit of
3.1\,$\times \,10^{-14}\, \rm erg \, cm^{-2} \, s^{-1}$ for the total exposures of {\fvv {26.8}} ks.
The X--ray ULs are reported in Table \ref{tab:tab6}.

\begin{table*}[ht!]
  \begin{center}
    \caption{\swiftp/{\fv {XRT}} 3-$\sigma$ flux ULs.}
    \begin{tabular}{|c|c|c|}

     \hline
     Start Date  & Effect.Exposure & 0.3--10\,keV UL value    \\
       (UTC)     & (s) & ($\times 10^{-13}\,\rm erg \, cm^{-2} \, s^{-1}$)\\
     \hline

     2020-02-03 18:43:55 & 1677 & 3.3 \\
     2020-02-04 15:43:59 & 1966 & 2.4 \\
     2020-02-05 15:23:58 & 1304 & 3.5 \\
     2020-02-06 15:29:51 & 2006 & 2.3 \\
     2020-02-07 15:23:35 & 2016 & 2.7 \\
     2020-02-08 15:16:35 & 1264 & 3.4 \\
     2020-02-25 14:00:01 & 920  & 5.4 \\
     2020-03-05 13:08:00 & 1828 & 3.4 \\ 
     2020-03-06 14:23:00 & 1241 & 3.9 \\
     2020-03-07 14:16:00 & 1392 & 3.3 \\
     2020-03-08 14:12:13 & 303  & 15.0 \\
     2020-03-09 14:04:46 & 1597 & 2.9 \\
     2020-03-10 13:57:52 & 920  & 5.4 \\
     2020-03-23 12:20:41 & 1795 & 3.5 \\
     2020-03-24 12:40:46 & 727  & 7.2 \\
     2020-03-25 12:40:34 & 318  & 18.0 \\
     2020-03-26 12:20:14 & 1938 & 2.3 \\
     2020-03-27 11:53:55 & 1738 & 2.8 \\
     2020-03-28 11:48:08 & 1254 & 3.7 \\

     \hline

    \end{tabular}
    \label{tab:tab6}
 \end{center}
\end{table*}

\begin{figure*}[tbp]
\vskip-0.9cm
\begin{center}

  \centerline{\includegraphics[angle=-90,width=0.65\textwidth]{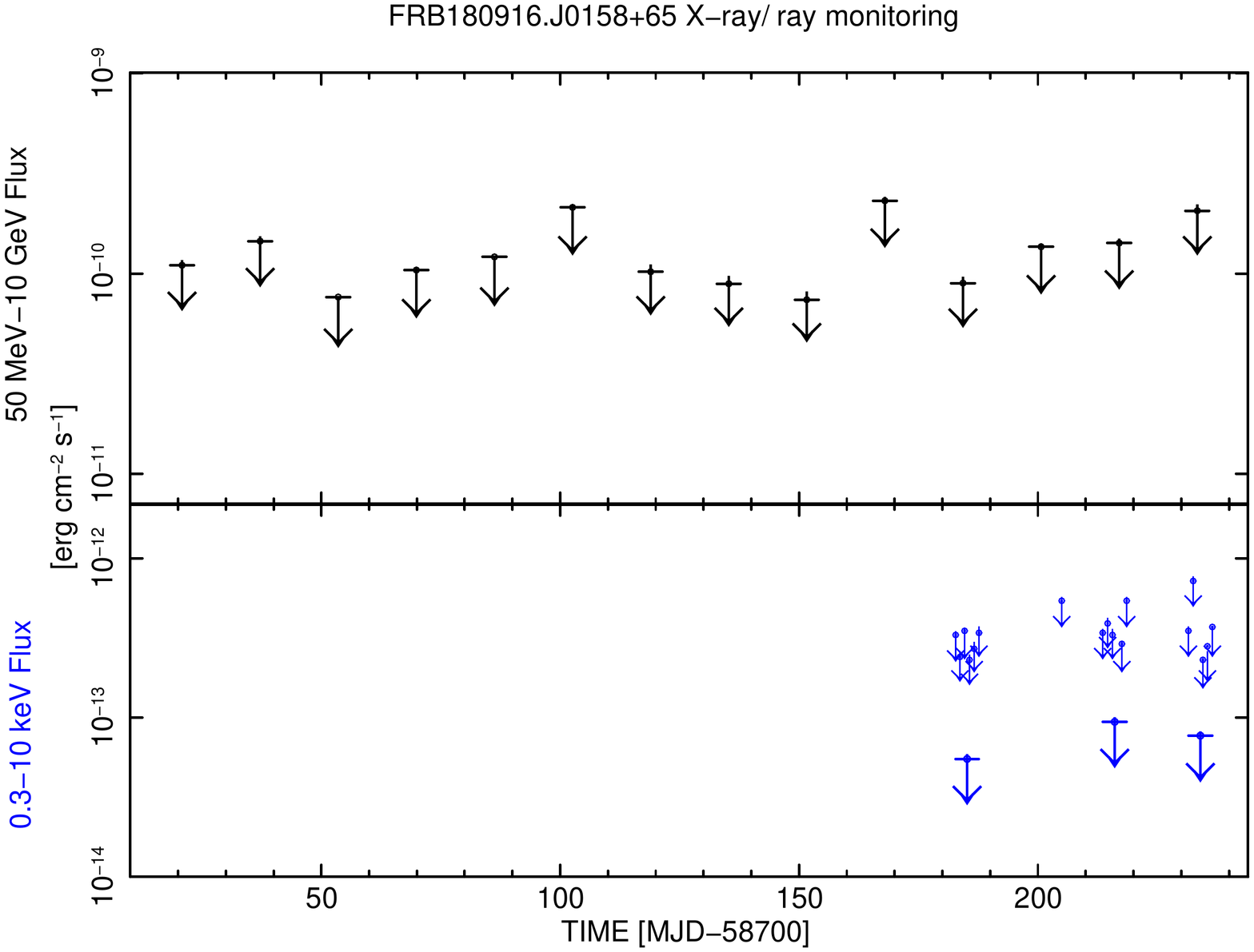}}
\vskip-1.0cm
  \centerline{\includegraphics[angle=-90,width=0.65\textwidth]{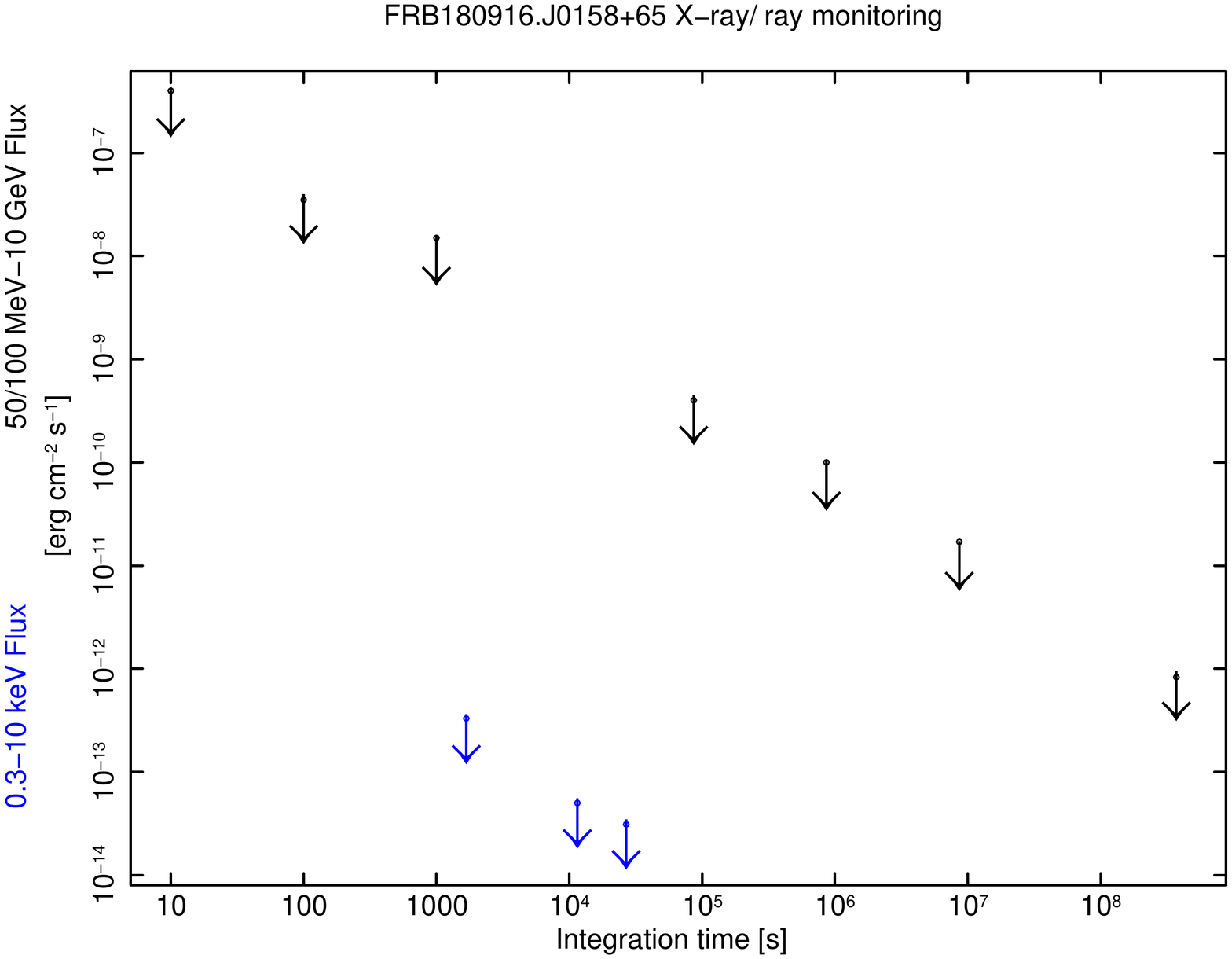}}
\vskip-0.3cm

  \caption{{\fv{(\textit{Top panel:})}} \textit{Upper plot: }\agilep/GRID 5-day {\mtt $\gamma$-ray flux ULs (100\,MeV - 10\,GeV)} obtained at {\mtt {active cycles of \frb during the period August 2019 - March 2020.}}
\textit{Bottom plot: }\swiftp/XRT flux ULs {\mtt marked in blue color}
 {\mtt (shown for integrations larger than 900 s)} in the 0.3\,--\,10 keV energy range;
{\mtt {they are obtained}} for individual observations during cycle 0{\fvv {, 2 and 3}} of {\cc {Table}} \ref{tab:tab6} ({\cc {Feb.\,3\,-\,8, Feb. 25, Mar.\,5\,-\,10 and Mar.\,23\,-\,28}}, 2020). The lowest values of the ULs
{\mtt {are for the summed observations of cycle 0 (10.2 ks), cycle 2 (7.9 ks) and cycle 3 (7.8 ks)}}.
{\fv{(\textit{Bottom panel:})}} {\mtt {AGILE and \swift}} flux ULs
obtained with different integration time intervals {\mtt of observations of \frbp.}
The GRID ULs (in black color) in the range 50\,MeV - 10\,GeV are for {\mtt integrations of 10, 100 and 1000 s centered on radio bursts of \frbp}, and in the range 100\,MeV - 10\,GeV for integrations of 1, 2, 100 days. The {\mtt lowest value} UL is obtained for a {\mt deep} GRID integration of $\sim$\,11.5 years.
 \swiftp/XRT ULs in the 0.3\,--\,10 keV energy range (in blue color) are
  {\mtt {shown for typical single 1.5 ks integrations, summed integration during an active cycle (cycle 0), and for the sum of all observations during cycles 0-3 with total exposure of 26.8 ks.}}
   }
  \label{fig2}
\end{center}
\end{figure*}





\newpage

\section{Discussion}

Our X-ray and $\gamma$-ray observations of \frb provide valuable
constraints on the nature of the source. The
(isotropic equivalent) energy of radio bursts is
{\mt $E_{radio, iso}\, \simeq \,(3 \times\, 10^{37}\, {\rm erg}) \, S_{\nu,
Jy} \, \delta t_{ms} \,
\Delta \nu_{GHz} \, d_{150M}^2$}, where the measured radio flux
density is $S_{\nu,Jy}$ in units of Jansky,
$\delta t_{ms}$ is the temporal width in units of milliseconds,
$\Delta \nu_{GHz}$ is the radio bandwidth in
units of GHz, and $d_{\rm 150Mpc} = d/{\rm 150 \, Mpc}$ with $d$ the
source distance from Earth. Note from Table 1 that the radio bursts of
\frb have radio spectral fluences $F'_{GHz}$ in the range $ 0.1 \,
{\rm Jy \, ms} \lesssim F'_{GHz} \lesssim 10 \, {\rm Jy \, ms}$ {\mtt for frequencies
in the interval $0.3 - 1.5 \rm \, GHz$}. We
therefore have $E_{radio, iso} \simeq  (10^{36}\, $--$\,  10^{38}) \, {\rm
erg}$ that is radiated on millisecond timescales. These energies are 6
orders of magnitude larger than those of the giant pulses from the
Crab pulsar \citep[][]{2003Natur422...141H}. It is  therefore unlikely that the underlying
source of \frb is a neutron star of the type associated with radio
pulsars in our Galaxy. Alternative possibilities to provide the required
  energies rely on compact objects,
either strongly magnetized objects such as magnetars, or black holes
(BHs). In both cases, the energetics involved in the impulsive radio
emission might be
a small fraction of a much larger latent power. The process of radio burst emission in
\frb is 
{\mt probably} "triggered" by the presence of a companion star.
  We can envision different possibilities for the generation of radio
pulses induced by a wind or plasma flow originating from a companion
{\mt (main sequence, O-B or Be)} star: magnetospheric events, accretion events, or shock interactions
with  mass outflows under the form of isotropic or equatorial winds as
in the case of Be stars. {\mt Alternative scenarios include periodicities
induced by a compact star precession or localized "spots" in accretion disks.}
{\mtt In case of an aliased period shorter than the 16.35 day value, precession or
spin of a magnetar-like object should be considered in addition to the binary hypothesis.}

We notice that no super-intense radio pulses of the type of \frb have
been detected from binary systems in our Galaxy. Probably the Galactic
compact objects that might generate intense radio pulses do not have
the right conditions to be "triggered" or induced in a proper way. For
example, Galactic magnetar-like objects might miss the right
combination of magnetic field energy and external "trigger" to be
effective as FRB sources.
Furthermore, no accreting BH in the Galaxy has ever been detected
producing short millisecond radio bursts such as those of FRBs. Again,
this might be the consequence of external conditions not easily
applicable to Galactic systems.
We are therefore dealing with an "exotic source" capable of
transcending the properties of known compact objects and their
environments, either Galactic neutron stars or black holes.
{\mtt Note, however, that rare Galactic radio bursts generated by FRB-like sources
might escape detection in searches by multiple-beam receivers as noticeably pointed
out by \citep[][]{2016ApJTendulkar}. In this case, the radio flux might be so intense as to saturate momentarily a receiver and be classified as radio frequency interference.
It is an interesting possibility for future searches to associate high-energy outbursts with simultaneous (within the delay times) very intense radio events classified as interference.}

Before speculating on the nature of the ultimate origin of \frbp, it
is useful to discuss the energetics.
At this moment, we obtained  upper limits to the high-energy emission
in 4 different energy bands from X-rays to $\gamma$-rays and for
different time intervals and integrations. We focus in this paper to
observations during the 5-day FRB-active intervals.

\textit{X-ray flux upper limits in the energy range 1-10 keV} (for
typical integrations of $10^3\, $s) as obtained by {\it Swift}
observations. The typical 1-2 ks UL is
$F_{X,UL} \sim 3 \times\, 10^{-13} \, \rm erg \, cm^{-2} \, s^{-1}$,
corresponding to an isotropic-equivalent X-ray luminosity $L_{X,UL}
\sim 8 \times\, 10^{41} \, \rm erg \, s^{-1}$. Summing the {\cc {7}} {\it
Swift} observations of {\cc {Table \ref{tab:tab6}}} we find an integrated UL
$F_{X,UL} \sim 5 \times\, 10^{-14} \, \rm erg \, cm^{-2} \, s^{-1}$,
corresponding to an isotropic X-ray luminosity $L_{X,UL} \sim 1.5
\times\, 10^{41} \, \rm erg \, s^{-1}$.

\textit{Hard X-ray upper limits in the range 18-60 keV} were obtained by Super-\agile during periods of favorable exposure in the time interval of this campaign. Summing up the observations, the UL is $1.8 \, \times 10^{-9} \, \rm erg \, cm^{-2} \, s^{-1}$.

\textit{MeV fluence upper limits in the range 0.4-1 MeV} (at the
millisecond level up to seconds) obtained by \agilep/MCAL simultaneous
with the FRBs of Table 1 marked with "YES" in the MCAL column. Typical
fluence UL in the sub- and millisecond range of integrations are
$F_{MeV,UL} \sim 10^{-8} \, \rm erg \, cm^{-2}$ corresponding to an
isotropic MeV energy UL $E_{MeV,UL} \sim 2 \times\, 10^{46} \, \rm erg
$.

\textit{GeV flux upper limits in the range 100 MeV -- 10 GeV} (for
integrations of hours/days/months) obtained by the \agilep/GRID during
active intervals of \frbp.
The typical 5-day flux UL is
$F_{\gamma,UL} \sim  10^{-10} \, \rm erg \, cm^{-2} \, s^{-1}$,
corresponding to an isotropic-equivalent $\gamma$-ray luminosity
$L_{\gamma,UL} \sim 3 \times\, 10^{44} \, \rm erg \, s^{-1}$. A more
stringent limit is obtained by summing all GRID exposures of the
\frb field over the interval 2007-2019: we find an integrated flux UL
$F_{\gamma,UL,ave} \sim 8 \times\, 10^{-13} \, \rm erg \, cm^{-2} \,
s^{-1}$, corresponding to an isotropic long-term averaged $\gamma$-ray
luminosity $L_{\gamma,UL,ave} \sim 2 \times\, 10^{42} \, \rm erg \,
s^{-1}$.

\subsection{The magnetized neutron star hypothesis}

Neutron stars of the magnetar-type are natural candidates for a class
of FRBs and should be considered for \frb {\mtt \citep[for a recent review on magnetars, see][]{2017ARA&A55...261}.}
Magnetic instabilities or sudden magnetospheric particle acceleration
phenomena can lead to short radio bursts. The action by an "external
driver" in a binary system can provide intruding particles and/or e.m.
fields that may lead to instabilities.
  The  magnetic fields associated with magnetars
(of the order of $B_m \sim 10^{14} - 10^{16} \, \rm G$) lead to
  maximal energies $E_m \sim R_m^3 B_m^2 / 6 \simeq (2\times\, 10^{49} erg)
  \, R^3_{m,6} \, B^{2}_{m,16}$, with $B_{m,16}$ the magnetar inner
magnetic field
  in units of $10^{16}\,\rm G$ and where we assumed a magnetospheric
radius of the magnetar
$R_{m,6} = R_m / (10^6 \, \rm cm)$.
A fraction of the total magnetic energy can therefore be dissipated 
  {\mt by} instabilities on timescales related to plasma relaxation.
  The ambipolar diffusion (dissipation) timescale of the inner
magnetar field has been estimated to be
  $\tau_d \sim 10^{11} \, \rm s$ \cite[][]{thompson1996} and leads to
an average dissipated luminosity
\be L_m \sim E_m / \tau_d \sim (10^{38} \, {\rm erg \, s^{-1}}) \,
R^3_{m,6} \, B^{2}_{m,16} \,
\tau_{d,11}^{-1} \label{eq-1}  \en
where
$\tau_{d,11}$ is the dissipation time in units of $10^{11} \, $s.
Larger/smaller values of $\tau_d^{-1}$ \cite[e.g.;][]{beloborodov17}
and/or of $B^{2}_{m,16}$ imply larger/smaller average luminosities.
The luminosity $L_m$ of Eq. \ref{eq-1} provides a reference value
for interpreting our X-ray and $\gamma$-ray observations of \frbp.

We see that if an instability mechanism is occasionally triggered in
the magnetar system,
in order to be detected in our observations we require
{\mtt that the object emits}  within
a millisecond
in the radio band an energy 10-1000 times larger that that deduced by
the  average luminosity $L_m$.
This is a demanding but  not impossible requirement for a highly
magnetized system
subject to sudden relaxation (e.g., the case of the soft $\gamma$-ray
repeater \sgrp; \citealt[][]{palmer2005}).

\frb might be then a magnetar occasionally subject to sudden
instabilities that dissipate in a
millisecond what normally would have been emitted on longer timescales
of order of seconds/minutes/hours.
Eq. \ref{eq-1} provides a reference luminosity for
optical/X-ray/$\gamma$-ray emission that should dissipate
most of the instability energy. The expected average high-energy flux
at the \frb distance is expected to be
\be F \simeq 4 \times\, 10^{-17}  \, \xi \, b^{-1} \, R^3_{m,6} \,
B^{2}_{m,16} \, \tau_{d,11}^{-1} \,
\rm erg \, cm^{-2} \, s^{-1} \label{eq-2} , \en
where $\xi$ is the conversion factor from magnetic to radiated
energies, and $b$ a beaming factor.
In order to be detectable in the optical/X-ray/$\gamma$-ray band we need
enhancements
with respect to the average dissipated luminosity $L_m$ by factors of
$10^3 - 10^4$.
These requirement is certainly demanding, but not impossible to
satisfy in a highly unstable system.

Our upper limits therefore constrain the high-energy dissipation
related to the active periods
  of \frbp.
These considerations based on isotropic emission can be modified to
account for beaming effects.
If beaming plays a role in the FRB phenomenon, our energy constraints
can be made even lower,
and approach hour/day luminosities of the
order of those of Eq. \ref{eq-1}.

We conclude that a magnetar with extreme properties in terms of its
magnetic field or dissipation process might power the \frb radio
bursts and be detectable in the X-ray range by current instruments.

\subsection{The black hole hypothesis}

A black hole interacting with a gaseous/plasma environment in a
binary system should be considered and cannot be discarded based on
our data.
In order to power the required detectable luminosities the mass of the
BH has to be
  in the range from above solar to hundreds or more of solar masses.
The radio and high-energy luminosities to be produced by the BH is
most likely accretion-powered and values larger that that of Eq.
\ref{eq-1} can be obtained for limited time intervals.

{\mtt Following \citep[][]{2020Natur577...190}, we remark} that the radio position of \frb is positionally coincident
with a remarkable "cusp" of star formation in its host galaxy.
If this is not a coincidence, the apparent cusp can
actually be a "wake" of stars caused by the presence of a BH of a mass
sufficiently large to influence an otherwise ordered flow.
Therefore, in principle the BH mass can be of order of thousands solar
masses or more for the presence of an intermediate mass BH or the
remnant of a galactic
nucleus as a consequence of a galaxy merging.

The BH hypothesis cannot at the moment be discarded, especially if we consider
the energetics of the bulk of FRBs that requires energies and
luminosities emitted in the radio (and presumably at higher energies) substantially larger
than those of \frb \citep[e.g.,][]{2019ARA&A119..161101}.

\section{Conclusions}
\label{concl}

\agile and \swift {\cc {joint}} observations of \frb set important constraints on the
energetics of the source by providing the best upper limits on the
X-ray and $\gamma$-ray emissions.

We focused in this paper on the 5-day time intervals during which \frb
is more active in producing radio bursts according to C20. A
multi-frequency campaign of \frb involved radio (SRT, Northern Cross), X-ray
(\swiftp) and $\gamma$-ray (\agilep) observations starting on {\cc {Feb.\,3, 2020 and
ending on Feb.\,8, 2020}}. In addition, we retrieved and studied \agile data for
several source active cycles during the time interval {\cc {Aug.\,25}}, 2019 -- {\fvv {Mar.\,28}}, 2020.
We also checked all {\fvv 35} known radio bursts from \frb searching
for any quasi-simultaneous transient detected by \agilep.

Our best upper limits can be used in the context of a magnetar-like object
dissipating its magnetic energy in a timescale $\tau_d$. If $B_m$ is
the equivalent
strength of the magnetic field that corresponds to the dissipated
portion of the magnetic energy,
our constraints translates into the relation (assuming $\xi \, b^{-1} \sim 1$)
\be  R^3_{m,6} \, B_{m,16}^2 \, \tau_{d,8}^{-1} \lesssim 1 , \label{eq-4} \en
  where  
$\tau_{d,8}$ is in units of $10^8$ {\cc {s}}.
Eq. \ref{eq-4} shows that magnetic dissipation (relaxation) timescales of the
order of our X-ray and $\gamma$-ray observations ($10^3 - 10^5$ {\cc {s}}) are
incompatible
with the presence of a very strong magnetic field of order of $10^{16} \, $G.
On the other hand, the very impulsive nature of the FRB phenomenon
implying millisecond
timescales is indicative of short durations for particle energization
and radio emission.
Our MeV upper limits obtained at millisecond timescales interpreted in
terms of isotropic emission
are close to the peak energy detected in the MeV range from \sgr
($2\,\times\, 10^{46} \rm \, erg$ emitted in about 200 {\cc {ms}}).
After an impulsive phase leading to the radio bursts,
the system may "relax" on a longer timescales with associated X-ray
and possibly $\gamma$-ray emission.
Eq.\ref{eq-4} is then an important (and best so far) constraint in
the context of a magnetar model for \frbp.

The search for high-energy counterparts of FRB sources continues,
and more data on \frb and other FRBs are necessary.
Considering the variability in the interaction between a companion
star and a compact
object near periastron as observed in X-ray binaries, a systematic
monitoring of \frb
is necessary.
\agile will then continue to monitor and search for high-energy emissions from \frb,
our so-far unique periodic-repeating FRB.

\vspace{2.0cm}

{\bf Acknowledgments:}
Investigation carried out with partial support by the ASI grant no. I/028/12/05. We would like to acknowledge the financial support of ASI
under contract to INAF: ASI 2014-049-R.0 dedicated to SSDC.
We thank B.Cenko and the entire Neil Gehrels Swift team for the help and support, especially the Science Planners and Duty Scientists B. Sbarufatti and J. A. Kennea for their help and professional support with the planning and execution of the ToO in the coordinated multiwavelength campaign. We thank V. Kaspi, E. Fonseca, and D. Li of the CHIME Collaboration for providing us the radio bursts topocentric arrival times.


\end{document}